\documentclass[twocolumn,
nofootinbib,
prd,aps,
tightenlines]{revtex4}
\usepackage{graphicx}

\def\nn{ \nonumber \\ }

\def\lqcd{\Lambda_{\text{QCD}}}
\def\mscale{\lqcd \sqrt{\lqcd/Q}}
\begin{document}

\title{Infrared Scales and Factorization in QCD}

\author{Aneesh V.~Manohar}
\affiliation{Department of Physics, University of California at San Diego,
La Jolla, CA 92093\vspace{4pt} }
\date{\today}
\begin{abstract}
Effective field theory methods are used to study factorization of the deep inelastic scattering cross-section. The cross-section is shown to factor in QCD, even though it does not factor in perturbation theory for some choices of the infrared regulator. Messenger modes are not required in soft-collinear effective theory for deep inelastic scattering as $x \to 1$.
\end{abstract}
\maketitle

Many high energy processes in QCD factor into the convolution of a perturbatively calculable short distance contribution, and a non-perturbative long distance hadronic function~\cite{book}. The classic example is deep inelastic scattering (DIS), where the structure functions factor into  short-distance hard-scattering kernels which depend on the large momentum scale $Q$, and parton distribution functions which depend on the non-perturbative scale $\lqcd$. DIS  as $x\to1$ has been studied using soft-collinear effective theory (SCET~\cite{Bauer:2000ew,Bauer:2000yr,Bauer:2001ct,Bauer:2001yt}), which allows one to systematically separate out the different momentum scales, and also sum the Sudakov double-logarithms~\cite{Manohar:2003vb,Chay:2005rz}.

Recently, it was argued~\cite{pecjak} that DIS does not factor as $x \to 1$ because of messenger modes~\cite{Becher:2003qh}  of momentum $\sim\mscale$, which cause the hadronic function to depend on $Q$ as well as $\lqcd$. The analysis of Ref.~\cite{pecjak} is based on a study of momentum regions contributing to perturbation theory graphs~\cite{regions}. The same method has been used in other examples, such as the calculation of the $B \to \pi$ form factor, where it was also claimed that messenger modes are important~\cite{Lange:2003pk}. 

The above claims are surprising. One would not expect the scale $\mscale$ to be relevant for hadronic structure functions and form factors. DIS has been extensively studied by diagrammatic methods, and has been shown to factor as $x \to 1$ into the convolution of a hard kernel, jet function, and parton distribution function, which depend on the scales $Q$, $Q \sqrt{1-x}$ and $\lqcd$, respectively~\cite{book}. The conclusions of Ref.~\cite{pecjak} contradict these results.

This paper discusses factorization and the (non) existence of messenger modes for DIS. It is shown that a naive evaluation of diagrams in perturbation theory with off-shell external states leads to the conclusion that messenger modes exist, and that structure functions do not factor, as found in Ref.~\cite{pecjak}. However, a confining gauge theory such as QCD does not require messenger modes, and factorization holds. The basic flaw in the argument is equating momentum regions that contribute to a Feynman diagram in perturbation theory with propagating modes in an effective field theory description of a confining gauge theory such as QCD. Infrared modes which exist in perturbation theory are not present due to confinement. In an effective theory such as SCET, perturbation theory is used to sum logarithms using the renormalization group, and to compute short distance matching conditions. The infrared part of both the full and effective theory amplitudes is not given by perturbation theory, but nevertheless, perturbative amplitudes can be used to compute the matching condition, which only depend on short distance physics~\cite{eft}. Perturbation theory is not used to compute the infrared dynamics of the theory. It is not necessary to include separate modes for infrared scales below $\lqcd$ that occur in perturbation theory diagrams, as is shown explicitly for DIS.

The role of the messenger scale in SCET will be clarified by studying a measurable
quantity, the spin-dependent ${g_1}_N$ structure function for DIS off a nucleon target. The structure function ${g_1}$ has been chosen rather than the unpolarized structure functions $F_{1,2}$ because all the computations needed for the analysis of $g_1$ have been given explicitly in the literature in the form needed here~\cite{am}. The conclusions derived below are general, and apply equally well to other problems studied using SCET.

The typical momentum scales in DIS at generic values of $x$ are $Q^2$ and $\lqcd^2$. The final hadronic state $X$ has invariant mass
\begin{eqnarray}
M_X^2 &=&{1-x \over x}  Q^2 + M_T^2
\end{eqnarray}
where $M_T$ is the target  mass. As $x \to 1$, 
 $M_X^2 \to Q^2(1-x) \ll Q^2$, so the final hadronic state $X$ becomes jet-like. DIS in the $x\to1$ limit can be studied~\cite{Manohar:2003vb,Chay:2005rz} using SCET. The large scale is $Q$, and the power-counting parameter $\lambda$ is taken to be $1-x=\lambda^2 \ll1 $. It is convenient to choose $\lambda^2 = \lqcd/Q$, so that the jet scale is $Q\lambda=\sqrt{Q \lqcd}$.

The standard factorization theorem for DIS~\cite{book} implies that
\begin{eqnarray}
{g_1}_N\left(x,{Q^2 \over \lqcd^2}\right)  &=& \int_x^1 {{\rm d}y\over y} \hat {g_1}_{q}\left(y,{Q^2 \over \mu^2}\right) f_{\Delta q/N}\left({x\over y},{\mu^2 \over \lqcd^2}\right)\nn
&&\hspace{-2cm} +\int_x^1{{\rm d}y\over y} \hat {g_1}_{G}\left(y,{Q^2 \over \mu^2}\right) f_{\Delta G/N}\left({x\over y},{\mu^2 \over \lqcd^2}\right)
\label{1}
\end{eqnarray}
where $\hat {g_1}_{q,G}$ are perturbatively calculable hard scattering kernels, and $f_{\Delta q/N}$, $f_{\Delta G/N}$ are the polarized quark and gluon distributions in the nucleon. The parton distribution functions depends on the target, but the hard scattering kernels do not. As $x \to 1$, the hard scattering kernels can be written as the convolution of a hard function with a jet function, and a soft function can be factored out of $f_{\Delta q, \Delta G}$~\cite{book,Chay:2005rz}. This additional factorization will not be important in what follows. The standard factorization form of Eq.~(\ref{1}) writes ${g_1}$  as a sum of terms of the form $K \otimes f$, where $K$ depends on $Q$ but not $\lqcd$, and $f$ depends on $\lqcd$ but not $Q$. ($\otimes$ denotes the convolution.) The main conclusion of Ref.~\cite{pecjak} is that the factorization form Eq.~(\ref{1}) needs to be modified so that the parton distribution functions also depend on $\mscale$, and hence on $Q$, so that even though the convolution form Eq.~(\ref{1}) holds, the $Q$ and $\lqcd$ dependence has not been factorized.

\begin{figure}
\begin{center}
\includegraphics[width=3cm]{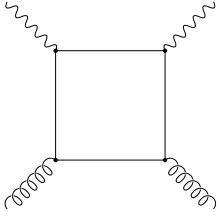}
\vspace{-0.5cm}
\end{center}
\caption{Box diagram contributing to deep-inelastic photon-gluon scattering via the process $\gamma^* + g \to q + \bar q$. The total cross-section is given by the imaginary part of the diagram. The wavy lines are the high-$Q^2$ virtual photon, and the helical lines are the gluon target.\label{fig:1} }
\end{figure}

The hard scattering kernels are target independent, and so can be computed from a perturbative computation of DIS off a partonic target. In particular, DIS off a gluon target can be written as
\begin{eqnarray}
{g_1}_G &=& \hat {g_1}_q \otimes f_{\Delta q/G} + \hat {g_1}_G \otimes f_{\Delta G/G}.
\label{2}
\end{eqnarray}
The graphs contributing to DIS scattering off a gluon target are the box graph in Fig.~\ref{fig:1} and related graphs given by permuting the external lines.
Evaluating the graphs explicitly in the $Q \to \infty$ limit for an external gluon state with $p^2 \not=0$ and a quark with mass $m$ and unit electric charge gives~\cite{zima,am}
\begin{eqnarray}
{g_1}_{G} &=& h{\alpha_s \over 4 \pi} \Biggl[\left(2x-1\right) \ln{Q^2(1-x) \over m^2 x- p^2 x^2(1-x)}
 \nn
&&\hspace{1cm}+3-4x +{p^2 x(1-x) \over m^2-p^2 x (1-x)}\Biggr].
\label{3}
\end{eqnarray}
where $h$ is the helicity of the gluon. This result shows that, in the limit $x \to 1$, the graph depends on the scales $Q^2(1-x)$, $m^2$, and $p^2(1-x)$. The mass $m$ and $p$ are taken to be infrared quantities of order $\lqcd$, so $p^2(1-x)$ is the messenger scale $\mscale$ in the limit $x \to 1$.

Eq.~(\ref{3}) can be used to compute the hard scattering kernel $\hat {g_1}_G$.
In Eq.~(\ref{2}), $\hat {g_1}_G$, and $f_{\Delta q/G}$ are of order $\alpha_s$, so one only needs the order one expressions
\begin{eqnarray}
f_{\Delta G/G} = h\delta(1-x),\qquad \hat {g_1}_q &=& \frac 1 2 \delta (1-x).
\label{4}
\end{eqnarray}
Substituting in Eq.~(\ref{2}) gives
\begin{eqnarray}
{g_1}_G &=& \frac 1 2  f_{\Delta q/G} + h \,\hat {g_1}_G +\mathcal{O}\left(\alpha_s^2\right),
\label{5} 
\end{eqnarray}
so to obtain $\hat {g_1}_G$ requires a definition of $f_{\Delta q/G}$.

The parton distributions are defined as the matrix elements of light-cone correlation functions of quark and gluon fields~\cite{Collins:1981uw}. The polarized quark distribution is the matrix element of the bilocal operator~\cite{am}
\begin{eqnarray}
O_{\Delta q}(k^+) &=& {1 \over 4 \pi} \int {\rm d}z^- e^{-i  z^-  k^+} \bigl[
 \bar \psi(z^-)W(z^-) \gamma^+\gamma_5 \psi(0)  \nn
&&\qquad +   \bar \psi(0) W^\dagger(z^-) \gamma^+\gamma_5 \psi(z^-)\bigr]
\label{6}
\end{eqnarray}
where $\pm$ denote light-cone directions, and  $W(z^-)$ is a light-like Wilson line from $0$ to the point $y$ with light-cone coordinates $y^+=0$, $\mathbf{y}_\perp=0$, $y^-=z^-$.

\begin{figure}
\begin{center}
\includegraphics[width=3.5cm]{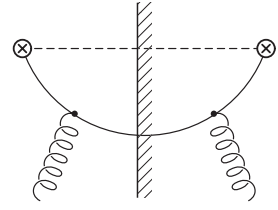}
\vspace{-0.3cm}
\end{center}
\caption{Graph contributing to the gluon matrix element of the quark distribution function. The two $\otimes$ and the dashed line denote the insertion of the bilocal operator $O_{\Delta q}$ defined in Eq.~(\ref{6}).  Graphs where the gluons connect to the Wilson line (the dashed line) vanish for the polarized parton distribution function.\label{fig:parton}}
\end{figure}

The parton distribution $f_{\Delta q/T}$ is the matrix element of $O_{\Delta q}$
in the target $T$ with momentum $P$,
\begin{eqnarray}
f_{\Delta q/T}(x) &=& 
\left\langle T,P \right | O_{ \Delta q} (x P^+) \left| T,P \right\rangle .
\label{6a}
\end{eqnarray}
 Using an off-shell gluon target in Eq.~(\ref{6a}) gives~\cite{am} (see Fig.~\ref{fig:parton})
\begin{eqnarray}
f_{\Delta q/G}&=&{\alpha_s \over 2 \pi} \Biggl[\left(2x-1\right) \ln{\mu^2 \over m^2 - p^2 x(1-x)}\nn
&&\hspace{1cm}-1+{m^2  \over m^2-p^2 x (1-x)}\Biggr],
\label{7}
\end{eqnarray}
which for $m=0$ reduces to
\begin{eqnarray}
f_{\Delta q/G}&=&h{\alpha_s \over 2 \pi} \Biggl[\left(2x-1\right) \ln{\mu^2 \over - p^2 x(1-x)}-1\Biggr].
\label{7z}
\end{eqnarray}
Substituting Eq.~(\ref{7}) in Eq.~(\ref{5}) gives
\begin{eqnarray}
\hat {g_1}_{G}&=&{\alpha_s \over 4 \pi} \left[\left(2x-1\right) \ln{Q^2(1-x) \over \mu^2 x}
+3-4x\right].
\label{8}
\end{eqnarray}

The computation of the kernel $\hat {g_1}_{G}$ is analogous to a matching computation in an effective field theory. One computes amplitudes in the full and effective theories,  each of which is infrared sensitive, in perturbation theory, and takes the difference to obtain the values of matching coefficients, which are independent of the infrared scales. Here, the structure function ${g_1}_G$ and the parton distribution function $f_{\Delta q/G}$ are both sensitive to the infrared regulators $m$ and $p$, but this dependence cancels in the kernel $\hat {g_1}_{G}$. In the limit $x \to 1$, 
\begin{eqnarray}
\hat {g_1}_{G}&\to&{\alpha_s \over 4 \pi} \left[ \ln{Q^2(1-x) \over \mu^2 }
-1\right],
\label{9}
\end{eqnarray}
and depends on the jet scale $Q^2(1-x)\sim Q \lqcd$, but not on any lower scales such as $\lqcd$ or $\mscale$, as expected. In SCET, the hard function is given by the matching coefficients at the scale $Q$, and the jet function is given by integrating out collinear modes at the scale $Q^2(1-x)$. The Sudakov double-logarithms as $x \to 1$ are summed by integrating SCET anomalous dimensions between $Q^2$ and $Q^2(1-x)$~\cite{Manohar:2003vb,Bauer:2000ew,Bauer:2003pi,drellyan}. All of these are perturbation theory computations, and are valid because the scales $Q^2$ and $Q^2(1-x)$ are both much larger than $\lqcd$.

That the hard scattering kernel  Eq.~(\ref{9}) is independent of infrared scales is usually taken as a proof of factorization. However, for factorization to hold, we also need the parton distributions to be independent of $Q$. An examination of the parton distribution Eq.~(\ref{7z}) for $m=0$ shows that when $ 1-x\sim \lqcd/Q$, $f_{\Delta q/G}$ depends on the messenger scale $\mscale$, and hence implicitly on $Q$, violating factorization. An evaluation of the diagram in Fig.~\ref{fig:parton} using the method of regions~\cite{regions} shows that the messenger region, with momentum of order  $p^2(1-x)$ contributes to the graph~\cite{pecjak}, and it is this region that lead to the $Q$ dependence in Eq.~(\ref{7z}). However, the parton distribution Eq.~(\ref{7}) for $m\not=0$ does not depend on the scale $p^2(1-x)$ for $x \to 1$, since in this limit $m^2 \gg p^2(1-x)$ and the logarithm depends only on $m^2$.

It is important to remember that neither the computation of ${g_1}_G$, nor that of $f_{\Delta q/G}$ is a valid computation. Both depend on infrared physics for which perturbation theory is not valid. Thus the dependence of ${g_1}_G$ and $f_{\Delta q/G}$ on $p^2(1-x)$, and the necessity of messenger modes to reproduce Eqs.~(\ref{3},\ref{7z}) is \emph{irrelevant}. What has been determined so far is that the kernel is given by Eq.~(\ref{8}), and this can be obtained by a perturbative effective field theory computation using SCET.  The dependence of the parton distribution on infrared scales is determined using Eq.~(\ref{6a}) for a \emph{nucleon} target. The relevant non-perturbative quantities it can depend on are the masses of physical hadrons that couple to the nucleon via the axial current matrix element. The spectrum of these hadrons cannot be computed in perturbation theory, but is known from both experimental observation and theoretical considerations to not contain states arbitrarily close in mass to the nucleon.  The nucleon parton distribution depends on the light-cone Fourier transform of the nucleon matrix element of the gauge invariant bilocal operator $\bar \psi(z^-) W(z^-) \gamma^+ \gamma_5 \psi(0)$. The $z^-$ dependence of this correlation function is governed by the scale $\lqcd$ in QCD, because of confinement, and so the parton distribution only depends on this scale.  Gauge invariance is crucial for the argument, since there can be massless gauge variant modes even in a confining gauge theory.  That the operator is bilocal, rather than local, is not important. Confinement implies there are no long range correlations even for non-local observables such as Wilson loops.

The above argument can be made more explicit by using the gauge invariant bilinears
\begin{eqnarray}
\Psi(z^-)&=& \phi_n^\dagger(z^-) \psi(z^-),
\end{eqnarray}
where $\phi_n$ is a color-triplet  field with Lagrange density $\phi^\dagger_n (i n \cdot D) \phi_n$. The sole purpose of $\phi_n$ is to reproduce the Wilson line $W(z^-)$, so there is no  longer an explicit Wilson line between $0$ and $z^-$.
In terms of these gauge invariant bilinears,
\begin{eqnarray}
O_{\Delta q}(k^+) &=& {1 \over 4 \pi} \int {\rm d}z^- e^{-i  z^-  k^+} \bigl[
 \bar \Psi(z^-) \gamma^+\gamma_5 \Psi(0)  \nn
&&\qquad +   \bar \Psi(0)  \gamma^+\gamma_5 \Psi(z^-)\bigr].
\label{6y}
\end{eqnarray}

Inserting a complete set of states $X$ between the $\Psi$ and $\bar\Psi$ fields, which is now possible because there is no Wilson line, shows that $O_{\Delta q}(x P^+)$ couples the nucleon to gauge invariant intermediate states with $P_X^+=P^+(1-x)$.  These intermediate states are color singlet hadrons made of two quarks and one color-triplet $\phi_n$ field. The $\phi_n$ field is equivalent to  a massless color-triplet quark moving in the light-cone $+$ direction. The parton distribution function depends on the spectrum of these excited states in QCD, which is governed by $\lqcd$. Thus the nucleon parton distribution function is not sensitive to scales below $\lqcd$. Note that there can be excited states with individual mass differences which are accidently much smaller than $\lqcd$. In the DIS regime, one sums over many final states, and what matters is the typical mass splitting, of order $\lqcd$. In inclusive $B$ decays, which are similar in structure to DIS, the relevant non-perturbative scale has been shown to be $\lqcd$~\cite{boyd}, even though there are individual states, such as the $B^*$, which have an order $\lqcd^2/m_b$ mass difference from the $B$.

It is instructive to see why factorization fails at the level of perturbation theory. The bilocal operator Eq.~(\ref{6}) is not Lorentz invariant, but is invariant under boosts along the $\hat z$-axis. It depends on the momentum $k^+$, and a renormalization scale $\mu$, but is independent of the large momentum $q^\mu$.  The matrix element of the bilocal operator can depend on the boost invariant combination $p^- p_X^+=p^2(1-x)$.
The perturbative computation of the parton distribution Eq.~(\ref{6}) depends on the spectrum of intermediate physical states in perturbation theory. Consider first the case of massless fermions, with $p^2\not=0$. This is the infrared regulator used in Ref.~\cite{pecjak} and in all the examples where messenger modes contribute. The intermediate states in Fig.~\ref{fig:parton} are $\phi_n \bar q$ states with massless quarks, and can have any invariant mass $\ge 0$. There are intermediate states with mass $p^2(1-x)=\mscale$, and so the parton distribution function depends on this scale, as can be seen from Eq.~(\ref{7}) with $m^2=0$. If $m^2 \not=0$, then the lowest intermediate state has invariant mass $m^2$. In the limit $x \to 1$, $m^2 -p^2x(1-x) \to m^2$, and the parton distribution is sensitive to the scale $m$, but not to the lower scale $p^2(1-x)$. The existence of the messenger scale $\mscale$ depends on the precise form of the infrared regulator~\cite{dorsten}. Messenger modes are needed if one wants to reproduce the DIS cross-section for an off-shell partonic gluon target with massless quarks. However, we are interested in the DIS cross-section off hadrons, not partons; the partonic cross-section was merely an intermediate step in the computation of the kernel $\hat {g_1}_G$. The existence of low-mass hadronic states is an artefact of perturbation theory. Low-mass states are not present in QCD, messenger modes are not needed, and factorization holds.

There is an alternate argument for messenger modes in DIS as $x \to 1$. The computation of DIS in the Breit frame shows that there are contributions from $\bar n$-collinear modes, $n$-collinear modes and ultrasoft modes. The $\bar n$-collinear mode describe the target, the $n$-collinear mode the final hadronic jet, and the ultrasoft modes  the low energy gluons. For the matching at the hard scale $Q$ and the running between $Q$ and the jet scale $Q^2(1-x) \sim Q \lqcd$, one can treat both the $n$ and $\bar n$ collinear modes as having an off-shellness (for the purposes of regulating the infrared divergences and power counting) $Q \lqcd$. In this case, the momentum of these modes in the Breit frame is $p_n^+ \sim \lqcd$, $p_n^- \sim Q$, $p_{\bar n}^+ \sim Q$, $p_{\bar n}^- \sim \lqcd$, and the ultrasoft gluons have typical momentum $p^2 \sim p_n^+ p_{\bar n}^- \sim \lqcd^2$. The target rest frame is obtained from the Breit frame by an order $Q/\lqcd$ boost along the collision axis. If, in addition, one lowers the off-shellness of the $\bar n$-collinear mode to $\lqcd^2$~\cite{Becher:2003qh}, the ultrasoft modes in the Breit frame get mapped into messenger modes in the target rest frame with $p^2 \sim \lqcd^3/Q$. These messenger modes are the analog of rescaled ultrasoft modes in the Breit frame which also have $p^2 \sim \lqcd^3/Q$ . However, an explicit computation shows that this is not the case~\cite{Manohar:2003vb}. The ultrasoft modes in the target rest frame reproduce the contributions of both the $\bar n$-collinear and messenger modes in the Breit frame. Equivalently, the parton distribution function in Eq.~(\ref{6}) involves only ultrasoft fields, and a direct computation of its matrix element~\cite{am} gives Eq.~(\ref{7}). Fig.~\ref{fig:parton} gets contributions from regions where some quark lines have invariant mass of order $p^2$, and others have invariant mass of order $p^2(1-x)$; however, both are included in the ultrasoft graph. 

In the target frame computation, the target quarks have an off-shellness of order $\lqcd^2$ from the very beginning, and are described by ultrasoft fields. Once the jet scale has been integrated out, the dynamics of these ultrasoft fields gives the non-perturbative parton distribution function. In the Breit frame computation, the $\bar n$-collinear and ultrasoft/messenger modes are both non-perturbative below the jet scale, and together reproduce the effects of the non-perturbative ultrasoft fields in the target rest frame. The non-perturbative effects of both Breit frame modes can be combined into a single non-perturbative function, which has no knowledge of the messenger scale. Modes such as the messenger mode arise if one tries to ``over-factor'' a cross-section, i.e.\ separate out all the different scales that appear in a perturbation theory graph into separate contributions, even if all the scales are non-perturbative. In DIS, they would appear if one tries to factor the non-perturbative parton distribution into separate non-perturbative pieces based on the momentum scaling of perturbation theory. 

Note that Ref.~\cite{Manohar:2003vb} did not attempt to compute the total cross-section for scattering off a partonic target; partonic matrix elements were used only to compute matching coefficients and anomalous dimensions. In these computations, for which perturbation theory is valid for all the dynamical modes, one can treat the $\bar n$-collinear and ultrasoft modes separately in the Breit frame. The matching conditions and anomalous dimensions are identical in the Breit and target rest frame. The two Breit frame modes below the jet scale, where they become non-perturbative, were combined together in a single non-perturbative function. 

It might be possible to factor a soft function out of the parton distribution $f = S \otimes \tilde f$, as suggested by the Breit frame calculation, where $S$ is the matrix element of Wilson line operators, and $\tilde f$ is the matrix element of $n$-collinear fields (see e.g.\ Ref.~\cite{Chay:2005rz}). The parton distribution $f$ renormalized at a low scale $\mu$ is independent of $Q$, and any definition of the split of $f$ into $S$ and $\tilde f$ in the theory below the jet scale, to be useful, should also be independent of $Q$, and so independent of the messenger scale.

SCET is an effective theory formulated with ultrasoft and collinear modes~\cite{Bauer:2000ew,Bauer:2000yr,Bauer:2001ct,Bauer:2001yt}, and provides an adequate description of DIS as $x\to1$ off {\sl hadronic} targets. It allows one to separate short distance contributions at the hard scale $Q^2$ and the jet scale $Q^2(1-x)$ from the non-perturbative contributions. It is not necessary to have separate propagating modes for every momentum scale that contributes to a perturbative diagram with some infrared regulator. In the target rest frame, for example, all non-perturbative effects are included in the ultrasoft contribution, and messenger degrees of freedom are not needed. In general, one does not need a propagating mode in an effective field theory for every momentum scale in the problem with some infrared regulator. A simple example is NRQED, with expansion parameter $\alpha$. The Hydrogen and Positronium energy levels can be computed to high order using NRQED, which has potential ($E\sim m \alpha^2, p \sim m \alpha$), soft ($E\sim m \alpha, p \sim m \alpha$) and ultrasoft  ($E\sim m \alpha^2, p \sim m \alpha^2$) modes~\cite{nrqed}. The shape function for Positronium decay, which is analogous to the structure function in DIS, has been computed explicitly using this theory~\cite{shape}. It is not necessary to introduce additional modes at the scale $m\alpha^4$ of the hyperfine splitting, even though the ortho-para Positronium mass splitting $\Delta m=7 m_e \alpha^4/12$ occurs in energy denominators of the form $\Delta m - E_\gamma$.

I would like to thank C.~Bauer, Z.~Ligeti, I.~Stewart and M.~Trott for comments on the manuscript. This work was supported in part by DOE grant DE-FG03-97ER40546 and a Alexander von Humboldt Foundation research award.

{}

\end{document}